\documentclass[12pt,preprint]{aastex}

\def\a{\boldsymbol{a}}
\def\bb{\boldsymbol{b}}
\def\c{\boldsymbol{c}}

\def\D{\boldsymbol{D}}

\def\f{\boldsymbol{f}} 
\def\F{\boldsymbol{F}}

\def\g{\boldsymbol{g}}
\def\M{\boldsymbol{M}}

\def\x{\boldsymbol{x}}
\def\y{\boldsymbol{y}}

\def\F{\boldsymbol{F}}
\def\g{\boldsymbol{g}}

\def\U{\boldsymbol{U}}
\def\V{\boldsymbol{V}}

\def\ONE{\boldsymbol{1}}
\def\1{\boldsymbol{1}}
\def\0{\boldsymbol{0}}

\def\lamb{\boldsymbol{\lambda}}

\def\PSI{\boldsymbol{\psi }}
\def\RHO{\boldsymbol{\rho }}

\def\f{\boldsymbol{f}}

\usepackage{amsmath}
\usepackage{graphicx}
\usepackage{multirow}
\usepackage{subfigure}
\usepackage{natbib} 
\usepackage{color} 
\newtheorem{definition}{Definition}
\newtheorem{theorem}{Theorem}

\RequirePackage{color}

\slugcomment{ASTR-D-18-00210R2}
\shorttitle{Short article title}
\shortauthors{Pan et al.} 

\begin{document}
\title{A New Probability-one Homotopy Method for Solving Minimum-Time Low-Thrust Orbital Transfer Problems}
\author{Binfeng Pan, Xun Pan and Siqi Zhang}
\affil{College of Astronautics, Northwestern Polytechnical University, Xi'an, Shaanxi, China, 710072} 

\begin{abstract}
Homotopy methods have been widely utilized to solve low-thrust orbital transfer problems, however, it is not guaranteed that the optimal solution can be obtained by the existing homotopy methods. In this paper, a new homotopy method is presented, by which the optimal solution can be found with probability one. Generalized sufficient conditions, which are derived from the parametrized Sard's theorem, are first developed. A new type of probability-one homotopy formulation, which is custom-designed for solving minimum-time low-thrust trajectory optimization problems and satisfies all these sufficient conditions, is then constructed. By tracking the continuous zero curve initiated by an initial problem with known solution, the optimal solution of the original problem is guaranteed to be solved with probability one. Numerical demonstrations in a three-dimensional time-optimal low-thrust orbital transfer problem with 43 revolutions is presented to illustrate the applications of the method. 
\end{abstract} 


\section{Introduction}
\par The use of low-thrust propulsion in a variety of space missions~\citep{Rayman,Kawaguchi,Kugelberg} has gained great attention in the space community, which allows a substantial reduction of propellant consumption in virtue of its high specific impulse compared to traditional chemical propulsion. However, solving low-thrust trajectory optimization problems is known to be highly challenging, the solution methods of which are usually categorized as direct methods and indirect methods~\citep{Betts1998}. Direct methods convert the optimal control problems into nonlinear programming problems by appropriate discretization~\citep{Hargraves2,Armellin}, which are straightforward and robust to accommodate complex conditions. However, the optimality of the obtained solutions is not guaranteed. Indirect methods convert the original problems to two-point boundary-value problems~(TPBVPs) according to the optimal control theory, the solutions of which are guaranteed to be at least local extremals~\citep{Kechichian,Zeng2014Fast,Zhang2015Low,Jiang2016Systematic}. The main disadvantages associated with TPBVP are that its convergence domain is narrow and its solution is extremely sensitive to the initial unknowns if a single-shooting method is utilized, especially for the low-thrust trajectory optimization problems with long flight duration and many revolutions. Although multiple-shooting techniques exist that can efficiently enhance the robustness of the indirect methods, however, the number of the unknown variables may increase significantly~\citep{Betts1998,Taheri2017Co}.

\par Homotopy methods, the principle of which is that a given problem is embedded into a family of problems parameterized by a homotopic parameter, and the optimal solution to the original problem is obtained by tracing the optimal solutions of the embedded problems~\citep{Watson2002}, have been widely applied to circumvent the above disadvantages of solving low-thrust trajectory optimization problems by single shooting indirect methods. References~\citep{Bertrand, Haberkorn2004Low, Gergaud2006Homotopy, Guo2012, Jiang2012, Zhang2015Low, Chen2016Optimality, Enhanced2016, Chi2017Homotopy,Zhao2017Target, Pan2018} have successfully utilized homotopy methods to solve minimum-fuel low-thrust orbital transfer problems, the optimal thrusts of which are discontinuous bang-bang controls. In these methods, the homotopic parameter is embedded into the performance index to provide continuous transition of optimal controls from the initial problem to the original one. It has been widely observed that the original fuel-optimal low-thrust trajectory optimization problems can be easily solved with probability one once the initial solutions are achieved~\citep{Guo2012, Jiang2012,Zhang2015Low, Chen2016Optimality, Chi2017Homotopy, Pan2018}. In contrast, homotopy methods for solving minimum-time low-thrust orbital transfer problems, whose optimal thrusts keep constant during the whole optimal trajectory, are still not satisfactorily developed. In Refs.~\citep{Caillau2003,Yue2010Indirect,Caillau2012Minimum}, the homotopy parameter is embedded into the thrust magnitude, and the minimum-time problem with sufficiently large thrust magnitude is taken as the initial problem for the homotopic approach. However, it is also not guaranteed that the optimal solution to the original problem can be obtained. \citep{Pan2016Double} presented a new double-homotopy method to construct {\it discontinuous} homotopy path which connects the initial and the original problem. However, the construction of discontinuous homotopy path is only valid under an assumption that multiple branches of homotopy path always exist at specific homotopic parameter, which may not be ensured for different occasions. Hence, the convergence to the optimal solution of the original problem by the double-homotopy method is still not guaranteed. Thus, homotopy methods, which construct {\it continuous} homotopy path to solve minimum-time low-thrust orbital transfer problems with probability one, are still unsettled.

\par In this paper, a new probability-one homotopy method is presented to solve minimum-time low-thrust orbital transfer problems. Parametrized Sard's theorem~\citep{Chow1978Finding} and Watson's sufficient conditions~\citep{Watson2002} are first revisited, which ensure the probability-one convergence of the homotopy methods. Generalized sufficient conditions are then derived, and a new probability-one homotopy formulation is custom-designed to satisfy all the prerequisites of the generalized sufficient conditions for the minimum-time low-thrust orbital transfer problems. Numerical solutions of a minimum-time low-thrust orbital transfer problem are provided to demonstrate the effectiveness of the proposed method, the initial ratio of thrust-to-weight of which is as small as $6.8 \times {10^{ - 5}}$.

\section{Minimum-Time Low-Thrust Orbital Transfer Problem Formulation}
\label{Section:Problemformulation}
\par Consider the three-dimensional point-mass equations of motion formulated by the modified equinoctial orbit elements (MEOE), which are singularity free for all trajectories with inclinations not equal to 180 deg and defined as follows~\citep{Walker1985, Broucke1972On} 
\begin{eqnarray}
P &=&  a (1-e^2)
\label{EQ:MEOE1} \\
e_x &=& e \cos (\Omega + \omega) \\
e_y &=& e \sin (\Omega + \omega) \\
h_x &=& \tan(\frac{i}{2}) \cos (\Omega) \\
h_y &=& \tan(\frac{i}{2}) \sin (\Omega) \\
L &=&  \Omega + \omega + f
\label{EQ:MEOE6}
\end{eqnarray} 
In Eqs.~(\ref{EQ:MEOE1}-\ref{EQ:MEOE6}), $[P, e_x, e_y, h_x, h_y, L]$ are the six modified equinoctial orbit elements where $P$ is the semi-latus rectum, $e_x$ and $e_y$ are elements that
describe the eccentricity, $h_x$ and $h_y$ are elements that describe the inclination, and $L$ is the true longitude~\citep{Graham2016Minimum}. Besides, $[a,e,i,\Omega,\omega,f] $ are the six classic orbit elements where $a$ is the semi-major axis, $e$ is the eccentricity, $i$ is the inclination of the orbital plane, $\Omega $ is the right ascension of the ascending node, $\omega$ is the argument of periapsis, and $f$ is the true anomaly.   

\par For better numerical conditioning, the semi-latus rectum $P$, the time $t$ and the mass $m$ are normalized by the $R_0$, $\sqrt {R_0/g_0}$ and $m_0$ respectively, where $R_0=6371.004\rm{km}$ is the radius of the Earth at the equator, $g_0=9.8 \rm{m}/\rm{s}^2$ is the standard acceleration of gravity at sea level, and $m_0$ is the initial mass of spacecraft. With some abuse of notation in this note, $P$, $t$ and $m$ are still used to denote the dimensionless semi-latus rectum, time and mass. Thus, the dimensionless  three-dimensional point-mass equations of motion in terms of MEOE can be expressed as~\citep{Gergaud2007} 
\begin{eqnarray}
\x^\prime &=& \bb + \frac{u T_{max}}{m m_0 g_0} \M \ONE_T
\label{EQ:NONDIX} \\
m^\prime&=& -\frac{u T_{max}}{m_0 \zeta}
\label{EQ:NONDIM}
\end{eqnarray}
where $\x = [P, e_x, e_y, h_x, h_y,L]^{T}$, $T_{max}$ is the maximal thrust magnitude, and $\zeta$ is a constant determined by 
\begin{equation}
\zeta = \frac{I_{sp}}{ \sqrt{R_0 / g^3_0}}
\end{equation}
where $I_{sp}$ is the specific impulse of the engine. The engine thrust magnitude $ T=u T_{max}$ ($0 \leq u\leq  1$) and the unit vector of the thrust direction $\ONE_T$ are the controls to be determined. The vector $\bb$ gives the time rate of change of the states due to gravity, which is expressed as follows 
\begin{equation}
\bb = \left[0,0,0,0,0,\frac{W^2}{ P^{3/2} } \right]^T
\end{equation}
and the matrix $\M$ determines how the states change due to the thrust acceleration vector, which is defined by
\begin{equation}
\M =\sqrt{ P}
\left[
	\begin{array}{ccc}
	0 & \frac{2 P }{W} & 0  \\
	\sin (L)    &  \cos (L) + \frac{e_x + \cos (L)}{W}& -\frac{Z e_y}{W}  \\
	-\cos (L)  & \sin (L) + \frac{e_y + \sin (L)}{W}& \frac{Z e_x}{W} \\
	0 & 0& \frac{C \cos (L)}{2 W}  \\
	0 & 0& \frac{C \sin (L)}{2 W}  \\
	0 & 0& \frac{Z}{ W}  \\
	\end{array}
\right]
\end{equation}
where $W$, $Z$ and $C$ are scalars defined to be
\begin{eqnarray}
W &=& 1+ e_x \cos (L) + e_y \sin (L) \\
Z &=& h_x \sin (L) - h_y \cos (L) \\
C &=& 1+ h_x^2 + h_y ^2
\end{eqnarray}

\par For the minimum-time orbital transfer problem considered in this paper, the performance index is defined as follows
\begin{equation}
J = \int_{t_0}^{t_f} 1 dt
\label{EQ:J_time}
\end{equation}
where $t_0$ is the initial time with given value, and $t_f$ is the terminal time to be determined. The initial conditions are given, which are $\x_0 = [P_0, e_{x0}, e_{y0}, h_{x0}, h_{y0},L_0]^{T}$, and the terminal conditions are specified as 
\begin{equation}
 \PSI (\x(t_f),t_f)= [P(t_f) - P^\ast _f,e_{x}(t_f) - e^\ast_{xf},e_{y}(t_f) - e^\ast_{yf},h_{x}(t_f) - h^\ast_{xf},h_{y}(t_f) - h^\ast_{yf} ]^T = 0
\label{EQ:TERMINAL_CONSTRAINTS2}
\end{equation}
where $P^\ast_f$, $e^\ast_{xf}$, $e^\ast_{yf}$, $h^\ast_{xf}$ and $h^\ast_{yf}$ are the predefined target orbit elements. The true longitude at the terminal time $L_f$ is free.

\par According to the optimal control theory~\citep{Bryson:75}, the Hamiltonian is constructed as
\begin{eqnarray}
H &=& -1 + \lamb_x^T \left(\bb + \frac{u T_{max}}{m m_0 g_0} \M \ONE_T \right) - \lambda_m \frac{u T_{max}}{m_0 \zeta} 
\label{EQ:HAMILTONIAN_ORBIT}
\end{eqnarray}
where $\lamb_x = [\lambda_P,\lambda_{e_x},\lambda_{e_y},\lambda_{h_x},\lambda_{h_y},\lambda_L]^T $ and $\lambda_m$ are the costates associated with $\x$ and $m$, respectively. The corresponding governing differential equations of costates are expressed as
\begin{eqnarray}
\lamb_x^\prime &=& -\left(\frac{\partial{H}}{\partial{\x}}\right)^T =  - \left(\frac{\partial \bb}{\partial\x}\right)^T\lamb_x
-  \frac{u T_{max}}{m m_0 g_0} \frac{\partial (\lamb_x^T\M\1_T) }{\partial\x}
\label{EQ:COSTATES_ORBIT1}\\
\lambda_m^\prime &=& -\frac{\partial{H}}{\partial{m}} = \frac {u T_{max}}{m^2 m_0 g_0} \lamb_x^T \M \ONE_T 
\label{EQ:COSTATES_m}
\end{eqnarray}

\par To maximize the Hamiltonian defined in Eq.~(\ref{EQ:HAMILTONIAN_ORBIT}), the optimal thrust direction should be along the direction of $\M^T\lamb$, which is known as primer vector theory~\citep{Lawden:63} and expressed as follows
\begin{equation}
\ONE_T^{\ast} = \frac{\M^T\lamb_x}{\parallel \M^T\lamb_x\| }
\label{EQ:ONET_ORBIT}
\end{equation}

\par Substituting Eq.~(\ref{EQ:ONET_ORBIT}) into Eq.~(\ref{EQ:HAMILTONIAN_ORBIT}) yields
\begin{eqnarray}
H &=& -1 + \lamb_x^T \bb  + u T_{max} \left( \frac{\parallel \lamb_x^T \M \|}{m m_0 g_0} - \frac{\lambda_m} {m_0 \zeta}\right) :=  H_0 + S u
\label{EQ:HAMILTONIAN_ORBIT2}
\end{eqnarray}
where $H_0$ is independent of the thrust $T$, and $S$ is the switching function, which is expressed as
\begin{equation}
S =T_{max}( \frac{\parallel \lamb_x^T \M \|}{m m_0 g_0} - \frac{\lambda_m} {m_0 \zeta}\label{S})
\end{equation}
The optimal thrust magnitude is given by 
\begin{equation}
 u^{\ast} = \begin{cases}
  1, & S>0 \\
  0, & S<0\\
  0\leq u\leq 1, & S\equiv 0
  \end{cases}
\end{equation}

\par Substituting $\1_T^{\ast}$ in Eq.~(\ref{EQ:ONET_ORBIT}) into Eq.~(\ref{EQ:COSTATES_m}) yields
\begin{equation}
\lambda_m^\prime = u T_{max} \frac{\parallel \M^T\lamb_x\|}{ m^2 m_0 g_0}
\end{equation} 
which reveals that $\lambda_m^\prime >0$. Since $m(t_f)$ is free for the minimum-time orbital transfer problem, the corresponding transversality condition is $\lambda_m(t_f)=0$. It follows that $\lambda_m(t)<0$ for  $t\in [0, t_f)$. Therefore $S>0$ by its definition in Eq.~(\ref{S}). Hence the optimal thrust is always at the upper bound, which is expressed as follows 
\begin{equation}
u^{\ast} =  1
\label{EQ:T_ORBIT}
\end{equation}
\par The corresponding transversality condition associated with the free final true longitude $\L(t_f)$ is given as
\begin{equation}
\lambda_L(t_f) =  0
\label{EQ:TRANSVERSALITY_Lf}
\end{equation}

\par For this terminal time free orbital transfer problem, the corresponding transversality condition is given as
\begin{equation}
H(t_f) = H_0(t_f) + S(t_f) = 0
\label{EQ:TRANSVERSALITY}
\end{equation}
In 2008, Ref.~\citep{Lu2008} concluded that
\begin{equation}
H_0(t_f) = -1 + \lamb(t_f)^T \bb_f = 0
\label{EQ:NEW_TRANSVERSALITY1}
\end{equation}
is always automatically satisfied, thus the transversality condition defined in Eq.~(\ref{EQ:TRANSVERSALITY}) can be further simplified as 
\begin{equation}
 S(t_f) = T_{max}( \frac{\parallel \lamb(t_f)^T \M(t_f) \|}{m(t_f) m_0 g_0} - \frac{\lambda_m(t_f)} {m_0 \zeta}) = 0
\label{EQ:TRANSVERSALITY1}
\end{equation} 
Ref.~\citep{Lu2008} also presented an easier replacement to the above transversality condition, which is given as 
\begin{equation}
\parallel \lamb(t_f) \parallel =1
\label{EQ:NEW_TRANSVERSALITY2}
\end{equation} 
Thus, the mass costate $\lambda_m$ needs not be explicitly computed in this problem. It should be emphasized that both Eq.~(\ref{EQ:NEW_TRANSVERSALITY1}) and Eq.~(\ref{EQ:NEW_TRANSVERSALITY2}) are directly taken from Ref.~\citep{Lu2008}, the detailed derivations of which are omitted in this paper for brevity.
 
\par The $5$ terminal conditions in Eq.~(\ref{EQ:TERMINAL_CONSTRAINTS2}) and the $2$ transversality conditions in Eq.~(\ref{EQ:TRANSVERSALITY_Lf}) and Eq.~(\ref{EQ:NEW_TRANSVERSALITY2}) constitute the $7$ necessary conditions for the $6$ unknowns $\lamb(t_0)$ and the terminal flight time $t_f$, which are expressed as
\begin{equation}
\f(\y) =  \0
\label{EQ:NECESSARY_CONDITION_ORBIT}
\end{equation}
where $\y=[\lamb(t_0), t_f ]$.
 
\section{Fundamentals of Probability-One Homotopy Methods}
\label{SEC:HOMOTOPY}
\begin{figure}
\centering
\includegraphics[width=0.5\textwidth]{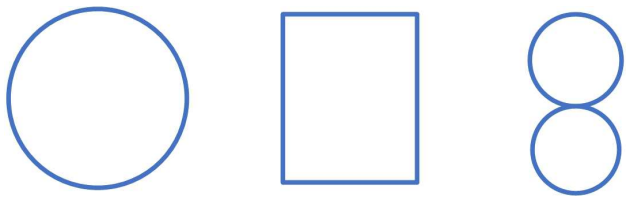}
 \label{Fig:2d2}
 \caption[]{Illustrations of topology homeomorphic properties}
 \label{Fig:homeomorphic}
\end{figure} 

\par Probability is the measure of the likelihood that an event will occur, which is quantified as a number between zero and one. If an event happens with probability one, it indicates that this event can happen almost surely. Although homotopy methods have been widely utilized to solve low-thrust orbital transfer problems, the probability-one convergence of which has not been well discussed. Thus in this section, fundamentals of the probability-one homotopy methods are provided as the basis for later discussion.

\subsection{Topology and Homotopy}
\begin{figure}
\centering
\subfigure[] { \includegraphics[width=0.3\textwidth]{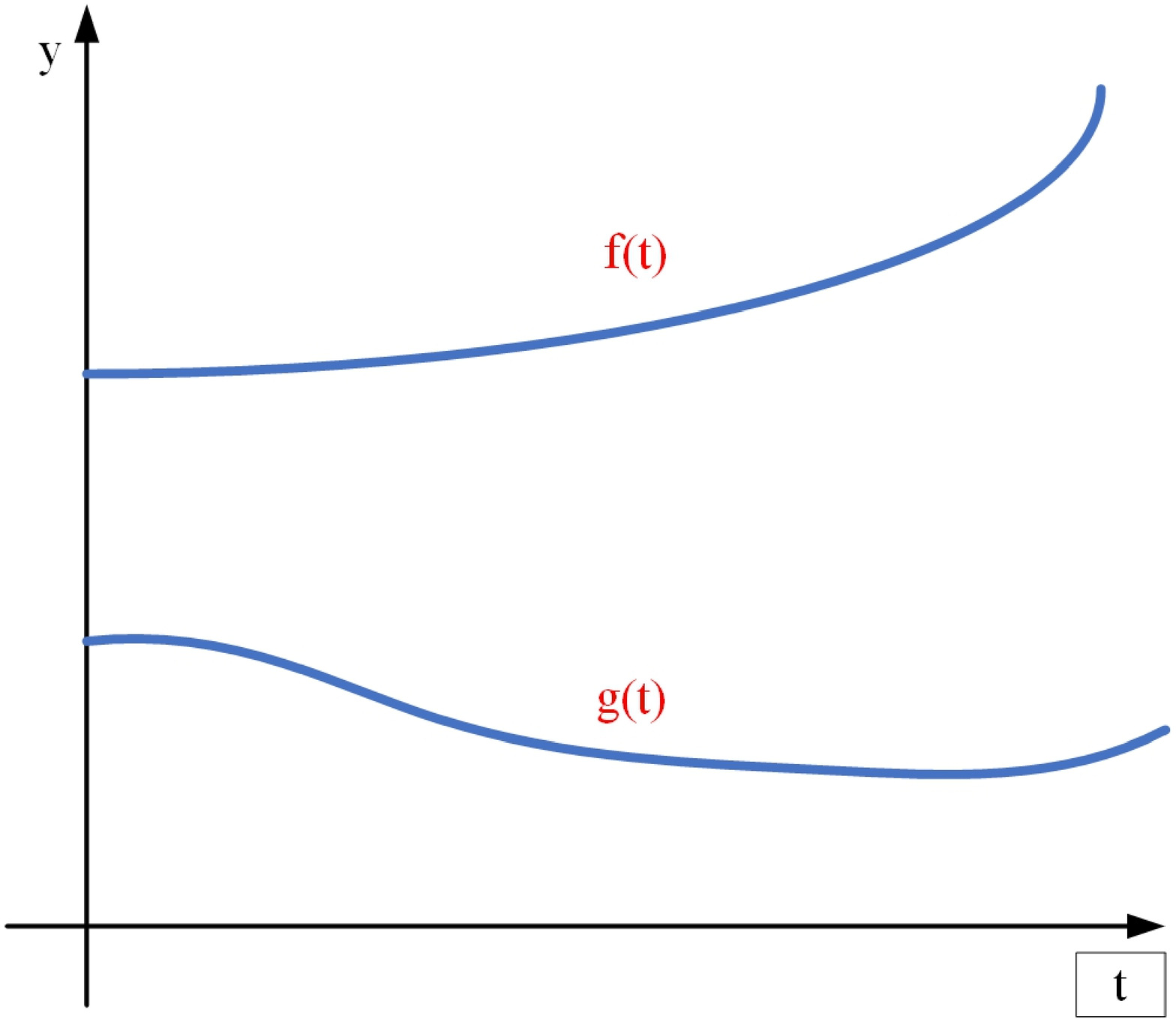}
 \label{Fig:h2d2}
 }
\subfigure[] { \includegraphics[width=0.3\textwidth]{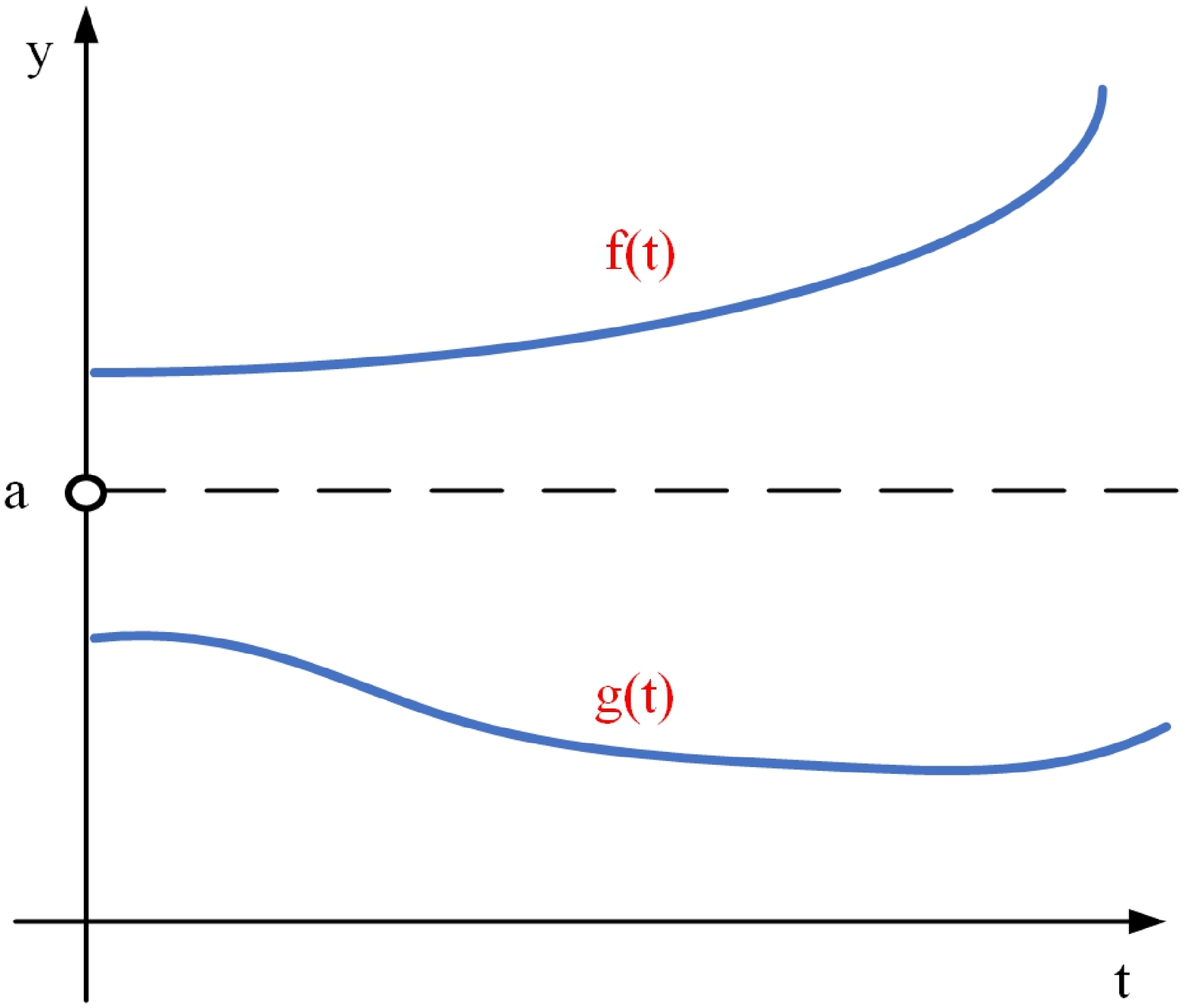}
 \label{Fig:h2d}
 }
 \caption[]{Illustrations of homotopic properties.}
 \label{Fig:homotopic}
\end{figure}  

\par In mathematics, {\it topology} is concerned with the properties of space that are preserved under continuous deformations, such as stretching, crumpling and bending, but not tearing or gluing\citep{Bourbaki1995General}. Intuitively, two spaces are {\it homeomorphic} if one can be deformed into the other without cutting or gluing. As illustrated in Fig.~\ref{Fig:homeomorphic}, a square and a circle are homeomorphic to each other,  but a figure 8 is not homeomorphic to a circle as the connection in the middle of a figure 8 has to be cut off. 

\par In topology, two continuous functions from one topological space to another are called {\it homotopic} if one can be continuously deformed into the other, such a deformation being called a \emph{homotopy} between the two functions\citep{Bourbaki1995General}. As illustrated in Fig.~\ref{Fig:h2d2}, a function $f(t)$ and a function $g(t)$ are homeomorphic to each other, but they are not if $y=a$ is not defined in the domain, as illustrated in Fig.~\ref{Fig:h2d}.  

\subsection{Basic Homotopy Methods}
\par  Consider the zero-finding problem of a system of nonlinear equations
\begin{equation}
\f(\y) = \0
\label{EQ:FX}
\end{equation}
where $\y\in R^n$, and $\f : R^n  \rightarrow  R^n$ is a smooth mapping, and a homotopy function is usually defined as an arbitrary smooth function, which is
\begin{equation}
\RHO (\kappa,\y), \quad \kappa\in[0, \ 1] 
\end{equation}
where $\RHO : R^n \times R \rightarrow  R^n$, and $\kappa$ is the homotopic parameter. This function continuously deforms a generally simpler function $\g(\y)$, which is chosen so as to ensure that its roots are known or easy to find at $\kappa=0$, into the original function $\f(\y)$ at $\kappa=1$. In this case, $\f$ and $\g$ are said to be {\it homotopic}. Typically, there are two types of homotopy methods, referred to as linear and nonlinear homotopy methods, respectively~\citep{Pan2016Double}.  In the linear homotopy methods, the homotopy function is a linear function of the homotopic parameter, which is defined as follows~\citep{Watson2002}
\begin{equation}
\RHO(\kappa, \y) = \kappa \f(\y) + (1-\kappa ) \g(\y)= \0, \quad \kappa\in[0, \ 1]
\label{EQ:LINEAR_HT}
\end{equation}
where $\g : R^n  \rightarrow  R^n$  is a smooth function having known solutions $\y_0\in R^n$ at $\kappa=0$. When the homotopic parameter $\kappa=0$, the homotopy function $\RHO (0,\y_0)= \g(\y_0)$, and when $\kappa$ is equal to $1$, the homotopy function $\RHO$ coincides with the original function $\f$. Please note that the solutions to $\RHO(\kappa ,\y)=\0$ usually have no physical interpretation for $\kappa <1$, and $\kappa=1$ is the value of interest. Typically, there are two most commonly used choices for $\g$, which can be identified by:
\par 1) It is called fixed-point homotopy method if $ \g(\y) = \y - \y_0$, which gradually deforms the function $\RHO(\kappa ,\y) $ from $\y=\y_0$ into $\f(\y)=\0$, and $\y_0$ is the unique solution of $\RHO(0, \y) = \0$, regardless of the structure of $\f(\y)$; 
\par 2) It is called Newton homotopy method if $ \g(\y) = \f(\y) - \f(\y_0)$, which gradually deforms the function $\RHO(\kappa,\y )$ from $\f(\y) - \f(\y_0)=0$ into $\f(\y)=\0$, and may have multiple roots for $\f(\y) = \f(\y_0)=\0$ at $\kappa = 0$.

\vspace{0.1in}
\par In nonlinear homotopy methods, the homotopy function is a nonlinear function of the homotopic parameter $\kappa$, which is defined as follows~\citep{Pan2016Double} 
\begin{equation}
\RHO(\kappa,\y ) =  \F(\kappa,\f(\y) ) = \0,  \quad \kappa\in[0, \ 1]
\label{EQ:NONLINEAR_HT}
\end{equation}
where $\F : R^n \times R \rightarrow  R^n$ is a nonlinear function of the homotopy parameter $\kappa$, which is carefully chosen such that $\RHO(1,\y) = \F(1,\f(\y) ) = \f(\y)$ and $\RHO(0,\y) =\F(0,\f(\y))=\g(\y_0)$.

\subsection{Probability-one Homotopy Methods}
In 1978, Chow {\it et al} first proposed a probability-one homotopy method to solve nonlinear equations\citep{Chow1978Finding}. The supporting theory is provided here for completeness and also as the basis for later discussion.

\begin{definition}~\citep{Chow1978Finding,Sielemann2012Probability}
Let $\U \subset R^n$ be open sets and $\RHO: \U \rightarrow R^p$ be smooth. We say $\RHO$ {\it is transversal to zero} if $\0 \in R^p$ is a regular value for $\RHO$.
\end{definition}

\begin{theorem}(Parametrized Sard's Theorem)~\citep{Chow1978Finding}
Let $\U \subset R^m$ and $\V \subset R^n$ be open sets, and let $\RHO: \U \times [0,1) \times \V \rightarrow R^n$ be a $C^2$ map. If $\RHO$ is transversal to zero, then for almost every $\c \in \U$ the map $\RHO_{\c}(\kappa,\cdot )= \RHO(\c,\kappa,\cdot )$ is also transversal to zero.
\label{THEOREM:PST}
\end{theorem}
\begin{figure}
 \centering\includegraphics[width=0.5\textwidth]{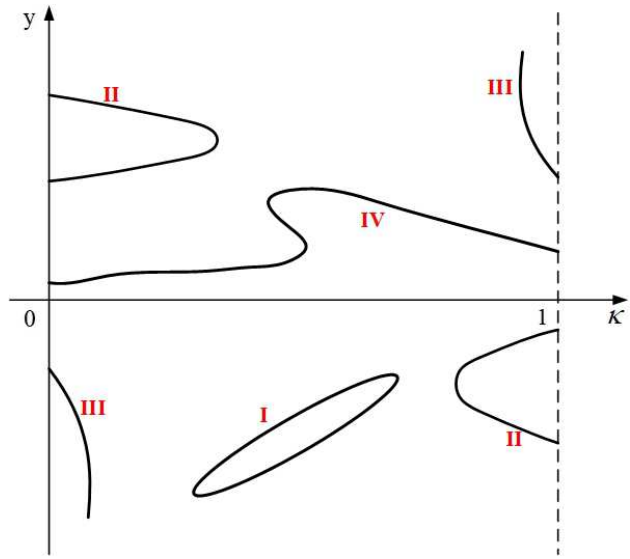}
 \caption[]{Possible curves of $\y(\kappa)$}
 \label{Fig:zerocurve} 
\end{figure}

\par In the above theorem, an additional parameter dependency on a random vector $\c \in R^m$ is introduced, and the $n\times( m+ n+1)$ dimensional Jacobian matrix $\D\RHO(\c,\kappa,\y)$ can be written as follows
\begin{equation}
\D\RHO(\c,\kappa,\y) = \left[ \frac{\partial \RHO }{\partial \c} \quad \frac{\partial \RHO }{\partial \kappa} \quad \frac{\partial \RHO }{\partial \y}   \right]
\end{equation}

\par Based on the Parametrized Sard's Theorem, the zero set of $\RHO_{\c}$ consists of several typical smooth, non-intersecting curves~\citep{Watson2002}, which are illustrated in $R^2$ as Fig.~\ref{Fig:zerocurve}:
\par 1) Type I: a closed loop entirely in $R^n \times  (0,1) $;
\par 2) Type II: a curve with both endpoints in $R^n \times (0)$ or $R^n \times (1)$;
\par 3) Type III: an unbounded curve with one endpoint in either $R^n \times (0) $ or $R^n \times (1) $,
\par 4) Type IV: a curve with one endpoint in $R^n\times (0) $ and the other in $R^n \times (1)$, which is called a {\it zero curve}.

\par Furthermore, for almost every $\c \in R^m $, the $n\times( m+ n+1)$ dimensional Jacobian matrix $\D\RHO(\c,\kappa,\y)$ has full rank at every point in $\RHO^{-1}_{\c}(0)=\{\y|\RHO(\c,\kappa,\y)=0\}$. Obviously, the goal of a homotopy method is to construct a zero curve with one endpoint in $R^n\times (0) $ and the other in $R^n \times (1)$. If a zero curve can be construct for almost every $\c \in R^m$, the corresponding homotopy methods are called {\it probability-one homotopy methods}~\citep{Chow1978Finding}. In other words, 
the probability-one homotopy methods can fail only for starting points in a set of Lebesgue measure zero. 

\par According to the parametrized Sard's theorem given in Theorem~\ref{THEOREM:PST}, Watson {\it et al}~\citep{Watson1987Algorithm} presented several sufficient conditions for probability-one homotopy methods with fixed-point homotopy type, which are summarized as follows:
\begin{theorem}(Watson's Sufficient Conditions)~\citep{Watson2002}
Let $\f: R^n \rightarrow  R^n$ be a $C^2$ map, $\RHO: \U \times [0,1) \times \V \rightarrow R^n$ a $C^2$ map, and $\RHO_{\c}(\kappa,\y)= \RHO(\c,\kappa,\y)$. Suppose that

(1) $\RHO$ is transversal to zero, 

and, for each fixed $\c \in R^m$, 

(2) $\RHO_{\c}(0,\y) =\0$ has a unique solution $\y_0$,

(3) $\RHO_{\c}(1,\y) =\f(\y)$ $(\y \in R^n)$. Then, for almost all $\c \in R^m$, there exists a zero curve $\gamma $ of $\RHO_{\c}$ emanating from $(0,\y_0)$, along which the Jacobian matrix $\D \RHO_{\c}$ has full rank. If, in addition,

(4) $\RHO^{-1}_{\c}(\0)$ is bounded, then $\gamma$ reaches a point $(1,\bar{\y} )$, where $\f(\bar{\y} ) =\0$. Furthermore, if $D \f(\bar{\y})$ is invertible, then $\gamma$ has finite arc length.
\label{THEOREM:PRO_ONE}
\end{theorem}

\par As mentioned previously, the prerequisites (1)-(4) of Theorem~\ref{THEOREM:PRO_ONE} are sufficient conditions, but not necessary. In order to apply this theorem, all the four prerequisites should be proved one by one. Homotopy maps can be easily constructed to meet prerequisites (2) and (3) by design. Prerequisite (1) may be trivial to verify for some homotopy maps and harder for others, in which $\kappa$ and $\a$ are involved nonlinearly. Prerequisite (4) is typically very hard to verify, and often is a deep result as (1)-(4) holding implies the existence of a solution to $\f(\y)=0$~\citep{Sielemann2012Probability}. That is why it has not been reported that Theorem~\ref{THEOREM:PRO_ONE} is utilized to construct a probability-one homotopy method for minimum-time low-thrust trajectory optimization problems.
\begin{figure}
	\centering
	\includegraphics[width=0.6\textwidth]{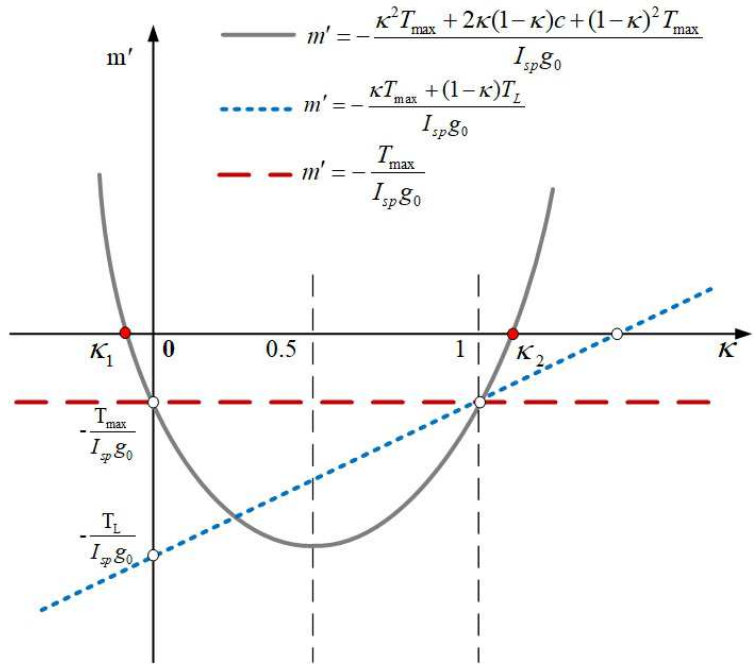}  
	\caption[]{Geometric illustrations of various mass rate functions with respect to the homotopic parameter}
	\label{Fig:p1}
\end{figure} 

\section{Probability-one Homotopy Method for Minimum-Time Low-Thrust Trajectory Optimization Problems}
\label{SEC:PROBABILITY}
\par In this section, a new probability-one homotopy method is presented to solve minimum-time low-thrust orbital transfer problems. A generalized parametrized Sard's theorem is first provided as supporting theory, which are:
\begin{theorem}(Generalized Parametrized Sard's Theorem)
Let $\U \subset R^m$ and $\V \subset R^n$ be open sets, and let $\RHO: \U \times [\alpha ,\beta) \times \V \rightarrow R^n$ be a $C^2$ map $(\alpha \leq 0,\beta \geq 1)$. If $\RHO$ is transversal to zero, then for almost every $\c \in \U$ the map $\RHO_{\c}(\kappa,\cdot )= \RHO(\c,\kappa,\cdot )$ is also transversal to zero.
\label{THEOREM:PST1}
\end{theorem}

\par The difference between Theorem~\ref{THEOREM:PST1} and Theorem~\ref{THEOREM:PST} only lies in the range of the homotopic parameter, which is $[0,1)$ in Theorem~\ref{THEOREM:PST} and $[\alpha ,\beta)$ $(\alpha \leq 0,\beta \geq 1)$ in Theorem~\ref{THEOREM:PST1}. According to Ref.~\citep{Chow1978Finding}, the parametrized Sard's theorem is actually valid for $\kappa \in (-\infty ,+\infty )$. Thus Theorem~\ref{THEOREM:PST1} is directly proposed according to Ref.~\citep{Chow1978Finding}, which is actually a generalized version of Theorem~\ref{THEOREM:PST}. Based on the above generalized parametrized Sard's theorem, Watson's sufficient conditions defined in Theorem~\ref{THEOREM:PRO_ONE} can be extended as follows:
\begin{theorem}(Generalized Watson's Sufficient Conditions)
Let $\f: R^n \rightarrow  R^n$ be a $C^2$ map, $\RHO: \U \times [\alpha ,\beta) \times \V \rightarrow R^n$ a $C^2$ map $(\alpha \leq 0,\beta \geq 1)$, and $\RHO_{\c}(\kappa,\y)= \RHO(\c,\kappa,\y)$. Suppose that

(1) $\RHO$ is transversal to zero, 

and, for each fixed $\c \in R^m$, 

(2) $\RHO_{\c}(0,\y) =\0$ may have several solutions,

(3) $\RHO_{\c}(1,\y) =\f(\y)$ $(\y \in R^n)$. Then, for almost all $\c \in R^m$, there exists a zero curve $\gamma $ of $\RHO_{\c}$ emanating from a starting point $(0,y_0)$, along which the Jacobian matrix $\D \RHO_{\c}$ has full rank. If, in addition,

(4) $\RHO^{-1}_{\c}(\0)$ is bounded, then $\gamma$ reaches a point $(1,\bar{\y} )$, where $\f(\bar{\y} ) =0$. Furthermore, if $D \f(\bar{\y})$ is invertible, then $\gamma$ has finite arc length.  
\label{THEOREM:PRO_ONE1}
\end{theorem}

\par Theorem~\ref{THEOREM:PRO_ONE} provides several sufficient conditions of probability-one homotopy methods, which is only valid for fixed-point homotopy. Thus in this paper, Theorem~\ref{THEOREM:PRO_ONE1} is presented, which can be extended to include nonlinear homotopy methods. In Theorem~\ref{THEOREM:PRO_ONE1}, the interval of homotopic parameter $\kappa$ is extended because of the situation that $\kappa$ beyond $[0,1)$ is also utilized in this new probability-one homotopy method. Besides, prerequisite (2) in Theorem \ref{THEOREM:PRO_ONE} that $\RHO_a(0,\y)=\0 $ has a unique solution is also relaxed to permit multiple solutions, the characteristic of which is also fully utilized in this proposed method. Thus it can be claimed that Theorem \ref{THEOREM:PRO_ONE1} is an extension of Theorem \ref{THEOREM:PRO_ONE}. Similar with Theorem~\ref{THEOREM:PRO_ONE}, homotopy maps can be easily constructed to meet prerequisites (1-3) in Theorem \ref{THEOREM:PRO_ONE1}, however, prerequisite (4) is still very hard to be verified.

\par In this paper, a new probability-one homotopy method is presented for solving minimum-time low-thrust orbital transfer problems, which satisfies all the 4 prerequisites in Theorem~\ref{THEOREM:PRO_ONE1}. This homotopy method is constructed by embedding the homotopic parameter $\kappa$ into the right-hand side of the equations of motion in Eq.~(\ref{EQ:NONDIM}), which is expressed as follows
\begin{eqnarray}
\x^\prime &=& \bb +   [\kappa T_{max}  + (1-\kappa)  T_L ] \frac{1}{m m_0 g_0 }\M \ONE_T 
\label{EQ:NONDIM_ORBIT3_x} \\
m^\prime&=& -\frac{\kappa^2 T_{max} + 2 \kappa (1-\kappa) c  + (1-\kappa )^2  T_{max}}{\zeta m_0}
\label{EQ:NONDIM_ORBIT3_m}
\end{eqnarray}
where the homotopic parameter $\kappa \in [\alpha ,\beta)$ $(\alpha \leq 0,\beta \geq 1)$, $T_L>>T_{max}$ is a sufficiently large thrust which makes it much easier to solve the low-thrust orbital transfer problems, and $c > T_{max}$ is a constant parameter. With the above equations of motion and  the performance index as in Eq.~(\ref{EQ:J_time}),  the Hamiltonian $H$ is rewritten as
\begin{equation}
H= -1 + \lamb_x^T \bb  + [\kappa T_{max}  + (1-\kappa)  T_L ] \frac{\lamb_x^T \M \ONE_T}{m m_0 g_0} - \lambda_m \frac{\kappa^2 T_{max} + 2 \kappa (1-\kappa) c  + (1-\kappa )^2  T_{max}}{\zeta m_0} 
\label{EQ:HAMILTONIAN_ORBIT3}
\end{equation}
and the corresponding  governing  differential equations for $\lamb_x$ are given as
\begin{eqnarray}
\lamb_x^\prime&=& -\left(\frac{\partial{H}}{\partial{\x}}\right)^T =  -\left(\frac{\partial  \bb}{\partial \x}\right)^T\lamb_x  - \left[ \frac{\kappa T_{max}  + (1-\kappa )  T_L }{ m m_0 g_0 } \right] \frac{\partial  (\lamb_x^T\M\1_T)}{\partial  \x}
\label{EQ:COSTATES_ORBIT2}
\end{eqnarray}
The optimal $\ONE_T$, $u$  and the necessary conditions remain the same as in Eqs.~(\ref{EQ:ONET_ORBIT}),   (\ref{EQ:T_ORBIT}), and (\ref{EQ:NECESSARY_CONDITION_ORBIT}).

\par In order to apply Theorem \ref{THEOREM:PRO_ONE1}, all the four prerequisites should be proved one by one:

\par 1) Prerequisite (1) in Theorem \ref{THEOREM:PRO_ONE1} indicates that at least one optimal solution exist at any homotopic parameter $\kappa$. For the minimum-time low-thrust orbital transfer problems, this phenomenon has been extensively observed in Refs.~\citep{Caillau2003,Yue2010Indirect,Caillau2012Minimum, Pan2016Double}. 

\par 2) Prerequisite (2) in Theorem \ref{THEOREM:PRO_ONE1} indicates that there may be multiple optimal solutions for the initial problem of the homotopy method, and prerequisite (3) in Theorem \ref{THEOREM:PRO_ONE1} requests that the embedding problem coincides with the original problem at $\kappa =1$, which are both satisfied naturally in this proposed method.

\par 3) Prerequisite (4) in Theorem \ref{THEOREM:PRO_ONE1} requests that the homotopic map of the embedding problems are bounded, which is guaranteed in this probability-one homotopy method by introducing a quadratic function of the homotopic parameter in Eq.~(\ref{EQ:NONDIM_ORBIT3_m}). 
A phenomenon has been observed in the literatures~\citep{Caillau2012Minimum,Pan2016Double,Pan2018_2}, that the singular point occurs when no solution exists with the specific revolution number, and then the homotopy curve turns backwards. Thus in this paper, the probability-one homotopy method with bounded homotopy curve is constructed by forcing the homotopy curve to turn backwards before it moves to the negative infinity.
The principle of the proposed method is simple and straightforward, a singular point is introduced into the negative homotopy branch by fictitiously increasing the spacecraft's mass along with the flight trajectory. As illustrated in Fig.~\ref{Fig:p1}, the mass rate turns to be positive when the homotopic parameter is beyond a specific boundary, which is determined by solving following equation 
\begin{equation}
\kappa^2 T_{max} + 2 \kappa (1-\kappa) c  + (1-\kappa )^2  T_{max} = 0
\end{equation}
the solutions of which are 
\begin{eqnarray}
\kappa_1 &=& \frac{c - T_{max} - \sqrt{c^2 -T_{max}^2 }}{2 (c - T_{max})} \\
\kappa_2 &=& \frac{c - T_{max} + \sqrt{c^2 -T_{max}^2 }}{2 (c - T_{max})} 
\end{eqnarray}
Once $\kappa < \kappa_1$ or $\kappa > \kappa_2$, the mass of the spacecraft grows rapidly along with flight time. Thus as long as the mass is sufficiently large, the homotopy path is forced to move backwards, which indicates that the homotopy path is bounded and prerequisite (4) in Theorem \ref{THEOREM:PRO_ONE1} is satisfied.  

\par It should be noted that the difference between this probability-one homotopy method and the ones utilized in Refs.~\citep{Caillau2003,Yue2010Indirect,Caillau2012Minimum, Pan2016Double} lies in the right-hand side of the differential equation of mass. In Refs.~\citep{Caillau2003,Yue2010Indirect}, the differential equation of mass is defined by
\begin{equation}
m^\prime = -\frac{\kappa T_{max}  + (1-\kappa)  T_L  }{I_{sp} g_0}
\end{equation}
and in Ref.~\citep{Pan2016Double}, the differential equation of mass is given as
\begin{equation}
m^\prime = -\frac{T_{max}}{I_{sp} g_0}
\end{equation} 
As illustrated in Fig.~\ref{Fig:p1}, the mass rate remains negative for both cases when $\kappa < 0$, thus it can be concluded that both homotopy paths are not bounded and the homotopy methods developed in Refs.~\citep{Caillau2003,Yue2010Indirect,Pan2016Double} are not probability-one methods. The simulation results provided in Refs.~\citep{Caillau2003,Yue2010Indirect,Pan2016Double} also validate the above conclusion.

\section{Numerical Demonstrations}
\label{SEC:NUMERICAL}
In this section, numerical results for a three-dimensional minimum-time low-thrust transfer problem are provided to demonstrate the effectiveness of the proposed probability-one homotopy method. The initial orbit is a geostationary transfer orbit (GTO), and the final orbit is a geostationary Earth orbit (GEO), the detailed conditions of which are summarized in Table~\ref{tab:orbit}. The initial mass of the spacecraft is 1500 kg, and the specific impulse is 2000 s. The maximum thrust magnitude is 1.0 N, which means that the initial thrust-to-weight ratio is only $6.8\times 10^{-5}$. Such a low thrust magnitude makes the problem very difficult to be solved. In this paper, all computations are executed on a desktop personal computer and all of the codes are implemented under Matlab. The absolute and relative error tolerances of the numerical integration algorithm are set as $10^{-14}$. The required accuracy in satisfying the final conditions of the TPBVP is set as $10^{-12}$.
\begin{table*}
\caption{Initial and final conditions in the low-thrust orbital transfer problem}
 \begin{center}
    \begin{tabular}{cccc}
      \hline
      $P_0$ (km) & 11623   & \qquad \qquad $P^\ast_f$ (km) & 42165\\
      $e_{x0}$  & 0.75   & \qquad \qquad $e^\ast_{xf}$ & 0.0 \\
      $e_{y0}$ & 0.0   & \qquad \qquad $e^\ast_{yf}$ & 0.0\\
      $h_{x0}$  & 0.0612   & \qquad \qquad $h^\ast_{xf}$ & 0.0\\
      $h_{y0}$  & 0.0   & \qquad \qquad $h^\ast_{yf}$ & 0.0\\
      $L_0 $(rad)  & $\pi$   & \qquad \qquad $L^\ast_f $(rad) &  free\\
    \hline
\end{tabular}
    \end{center}
  \label{tab:orbit}
\end{table*}

\par A much larger maximum thrust magnitude, $T_L = 30$ N, is used in the initial problem as defined in Eq.~(\ref{EQ:NONDIM_ORBIT3_x}) , which is much easier to be solved. According to Eq.~(\ref{EQ:NONDIM_ORBIT3_m}), the value of the constant parameter $c$ can be easily chosen as long as it is larger than $ T_{max}$, which is set to $20$ in this paper.

\par The initial problem with $\kappa = 0 $ can be easily solved by a simple single shooting method with the initial guesses of the costates $\lamb_0$ and the terminal time ${t}_{f}$ chosen as
\begin{eqnarray}
\widetilde{\lamb}_x(t_0) &=& [1,1,1,0,0,0]^T \\
\widetilde{t}_f &=& 100
\end{eqnarray}
Please note that these variables are all nondimensional quantities, among which the dimensional value of $t_f$ is 29.1858 hours. Thus the values of the converged costates and the terminal time are found to be:
\begin{eqnarray}
\lamb_x(t_0) & = & [0.4076615, 0.0103237, -0.0456599, -0.9049513, 0.1091981, 0.0275621]^T \\
 t_f & = & 130.3115
\end{eqnarray}

\begin{figure} 
	\centering
	\subfigure[The complete curve] { \includegraphics[width=0.46\textwidth]{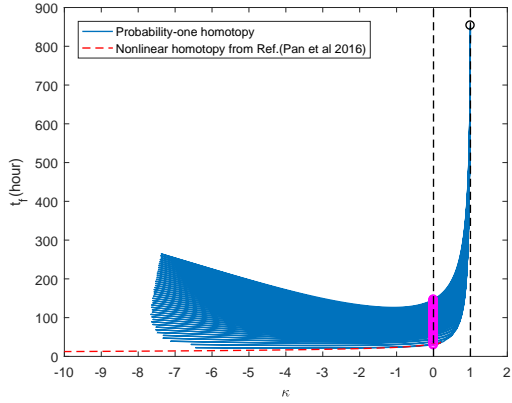}
		\label{Fig:tf2kappa_a}
	}
	\subfigure[The zoom-in view near $\kappa=0$] { \includegraphics[width=0.46\textwidth]{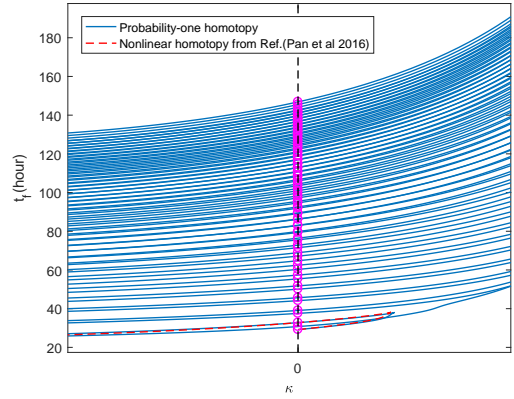}
		\label{Fig:tf2kappa_b}
	}
	\subfigure[The zoom-in view near $\kappa=1$] { \includegraphics[width=0.46\textwidth]{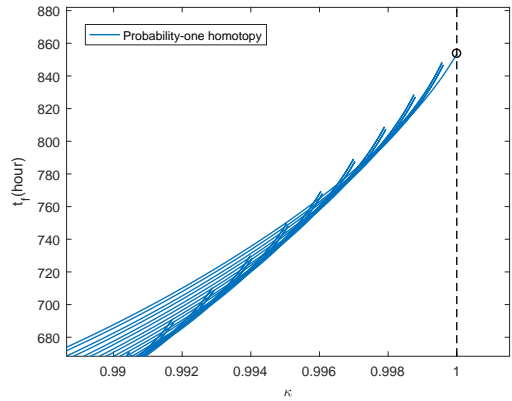}
		\label{Fig:tf2kappa_c}
	}
	\caption[]{Homotopy curves of $t_f$ by two different homotopy methods}
	\label{Fig:tf2kappa}
\end{figure} 
\begin{figure}
	\centering
	\subfigure[The complete curve] { \includegraphics[width=0.46\textwidth]{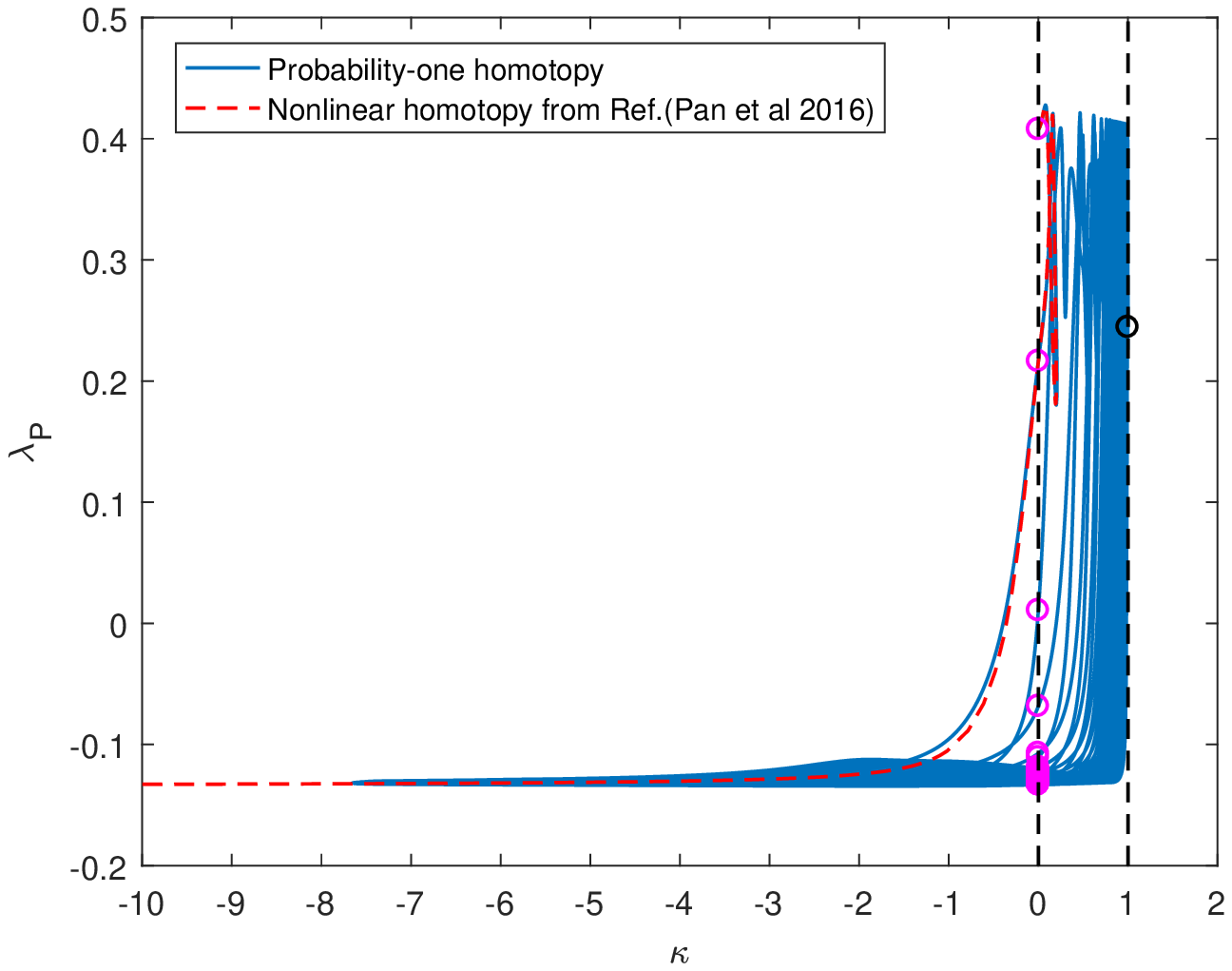}
		\label{Fig:lambdaP2kappa_a}
	}
	\subfigure[The zoom-in view of $0\leq \kappa \leq 1$] { \includegraphics[width=0.46\textwidth]{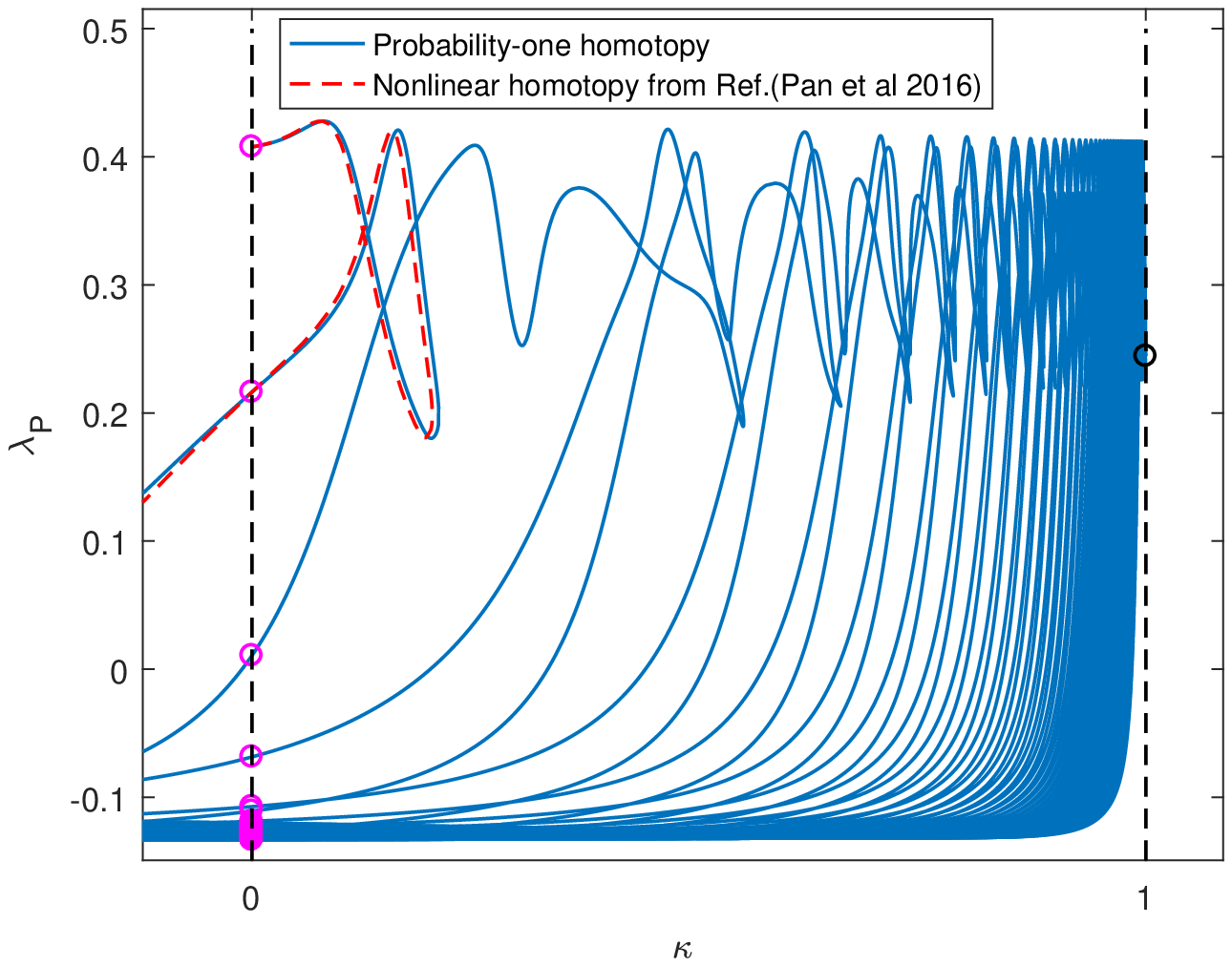}
		\label{Fig:lambdaP2kappa_b}
	}
	\caption[]{Homotopy curves of $\lambda_P(t_0)$ by two different homotopy methods}
	\label{Fig:lambdaP2kappa}
\end{figure} 
\begin{figure}
	\centering
	\subfigure[The complete curve] { \includegraphics[width=0.46\textwidth]{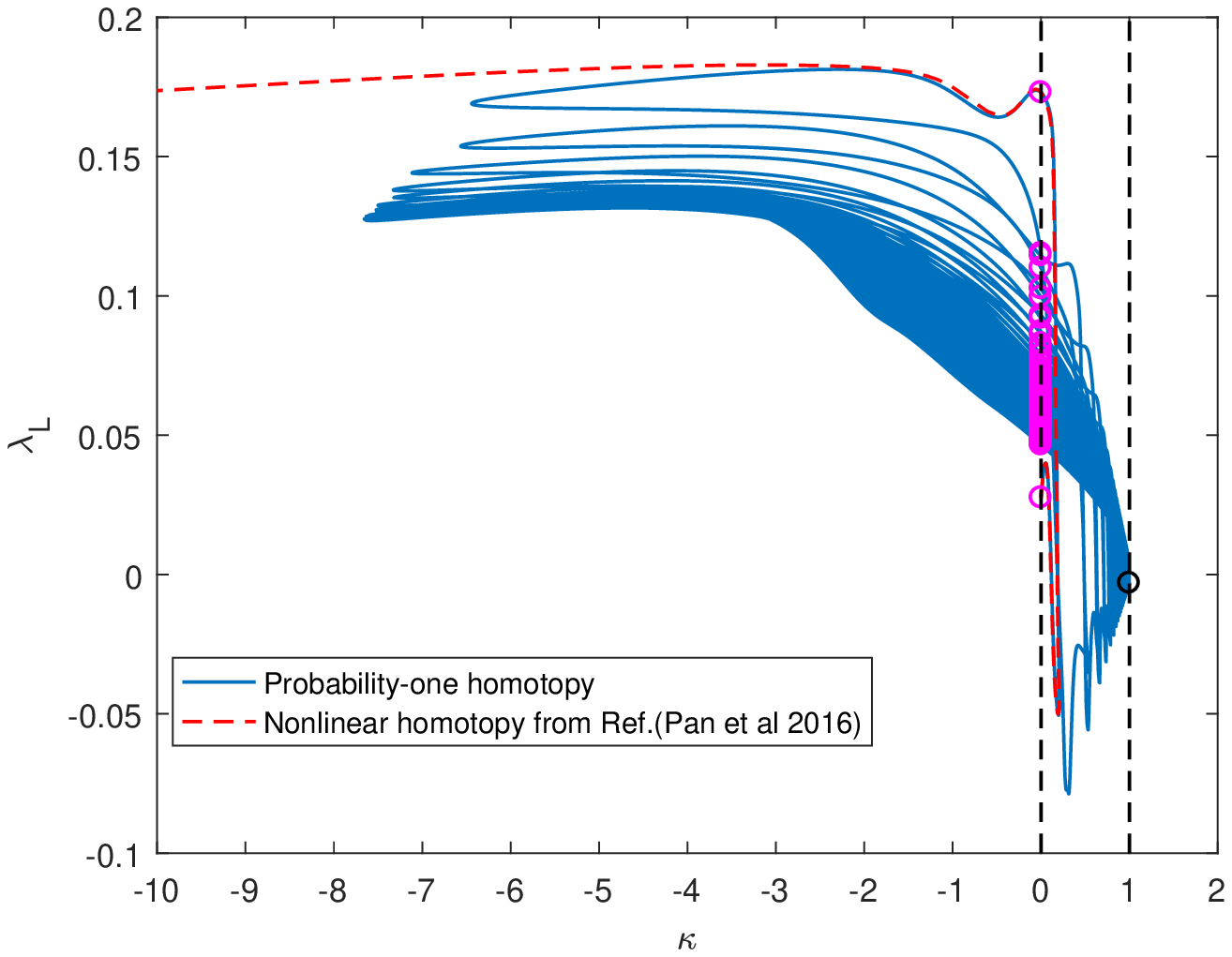}
		\label{Fig:lambdaL2kappa_a}
	}
	\subfigure[The zoom-in view of $0\leq \kappa \leq 1$] { \includegraphics[width=0.46\textwidth]{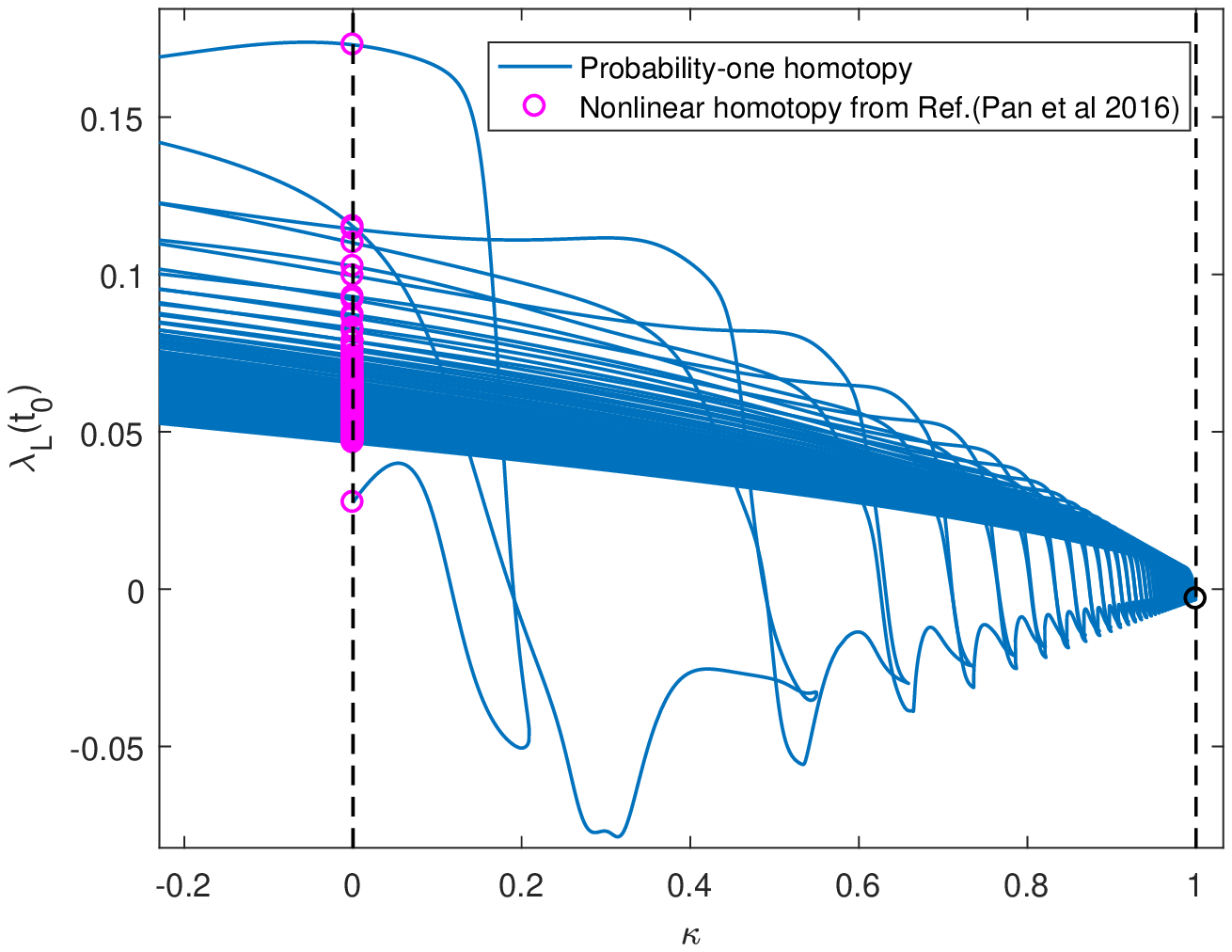}
		\label{Fig:lambdaL2kappa_b}
	}
	\caption[]{Homotopy curves of $\lambda_L(t_0)$ by two different homotopy methods}
	\label{Fig:lambdaL2kappa}
\end{figure} 
\begin{figure}
	\centering
	\subfigure[The complete curve] { \includegraphics[width=0.45\textwidth]{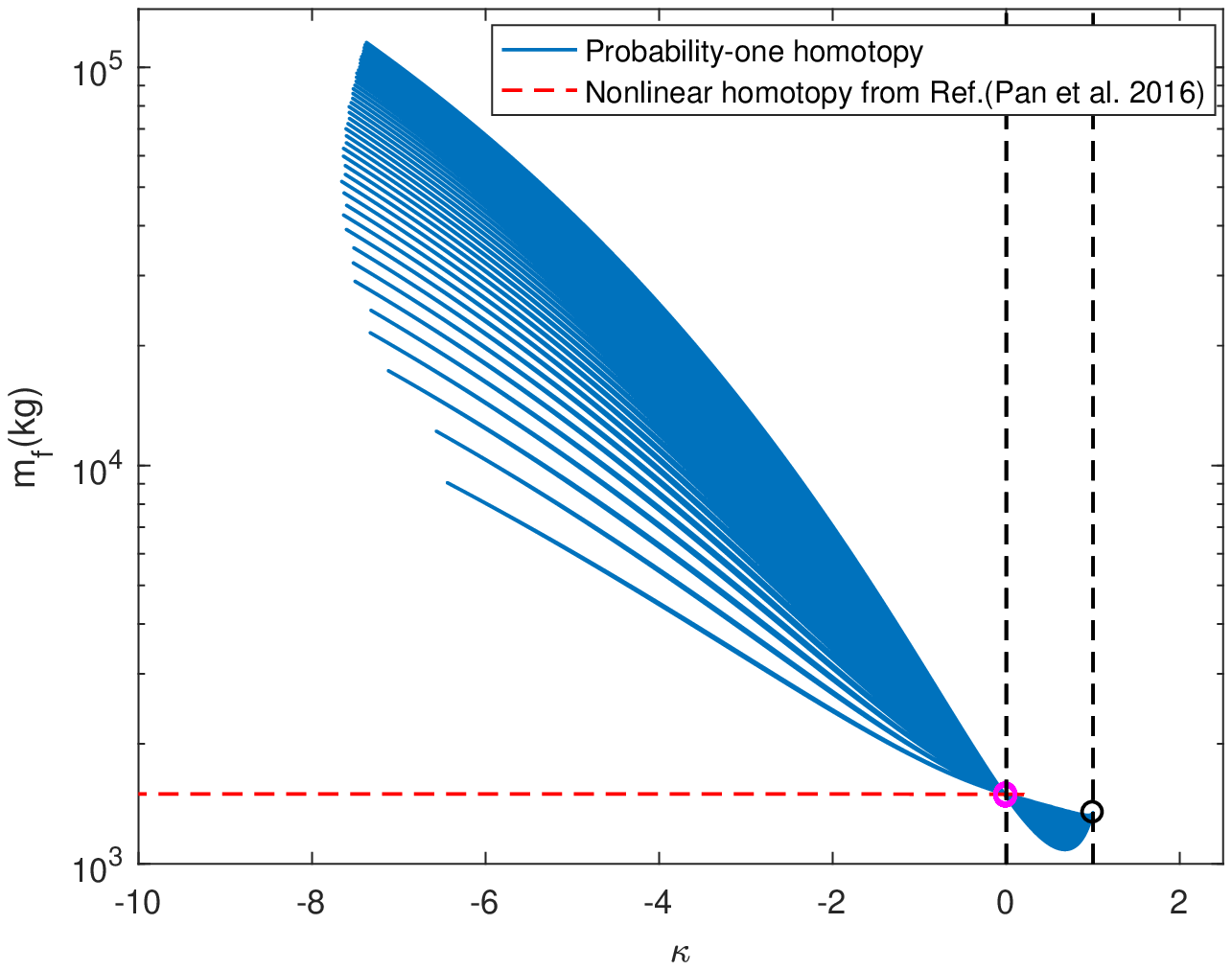}
		\label{Fig:mf2kappa_a}
	}
	\subfigure[The zoom-in view of  $0\leq \kappa \leq 1$] { \includegraphics[width=0.45\textwidth]{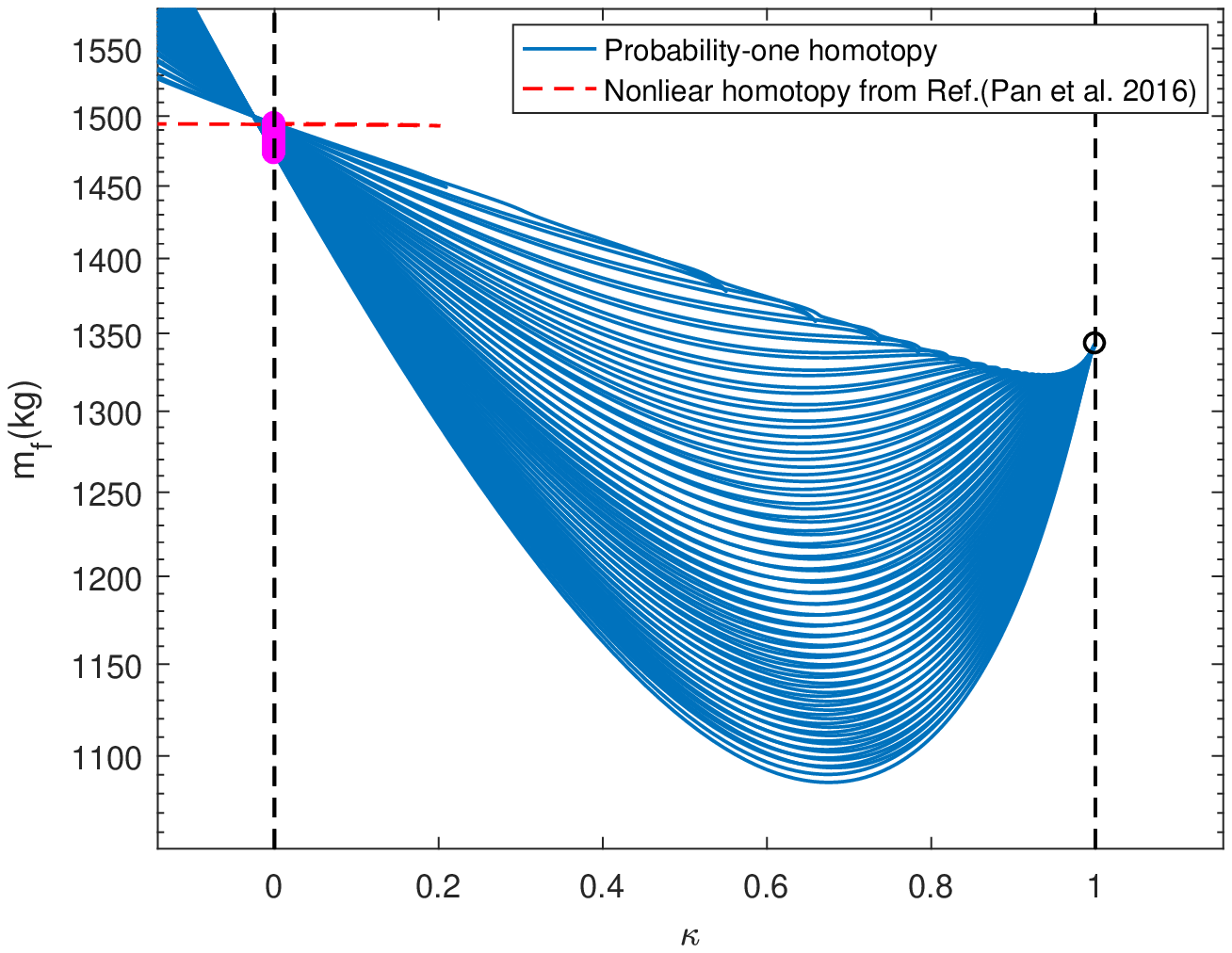}
		\label{Fig:mf2kappa_b}
	}
	\caption[]{Homotopy curves of $m_f$ by two different homotopy methods}  
	\label{Fig:mf2kappa} 
\end{figure} 
\par Initialized by this optimal solution of the initial problem at $\kappa =0.0$ as the starting point, the probability-one homotopy method is utilized to find the optimal solution of the original problem at $\kappa =1.0$. The pseudo-arclength continuation method~\citep{Keller1977} is used as path tracking algorithm to circumvent singular points in this paper, the initial correction step size of which is set to $\parallel \y \parallel/100$ in this paper. The details of the pseudo-arclength continuation method can be found in Refs.~\citep{Keller1977,Pan2016Double}. As illustrated in Figs.~\ref{Fig:tf2kappa}-\ref{Fig:mf2kappa}, the continuous zero curve is obtained by the proposed probability-one homotopy method, which finally reaches $\kappa =1$ after passing though multiple turning points. The algorithm is implemented on a desktop computer with a 2.20 GHz CPU, 4G RAM and Win10 operating system, which takes about 41 hours to obtain this solution in MATLAB. It should be noted that multiple solutions of the original problem at $\kappa = 0$ have been obtained, which are denoted by the pink circles in Figs.~\ref{Fig:tf2kappa}-\ref{Fig:mf2kappa}. In Figs.~\ref{Fig:tf2kappa}-\ref{Fig:mf2kappa}, a nonlinear homotopy method, which is taken from Ref.~\citep{Pan2016Double} is also applied for comparison. As illustrated in Figs.~\ref{Fig:tf2kappa}-\ref{Fig:lambdaL2kappa}, the homotopic parameter $\kappa$ first increases and the homotopy curve encounters a singular point at $\kappa=0.2018$, and then $\kappa$ decreases monotonically and approaches negative infinity, which fails to reach $\kappa=1$.
\par It should be emphasized that as illustrated in Fig.~\ref{Fig:mf2kappa}, the final mass of the spacecraft $m_f$ rises up to about $115000$ kg at around $\kappa = -7.3735$, which is $1500$ kg at $\kappa = -0.0256$. As described in Section \ref{SEC:PROBABILITY}, as long as the mass is sufficiently large, the homotopy path will be forced to move backwards, which guarantees the success of the probability-one homotopy method.

\par
\begin{figure}
	\centering
	\subfigure[The 1st solution] { \includegraphics[width=0.45\textwidth]{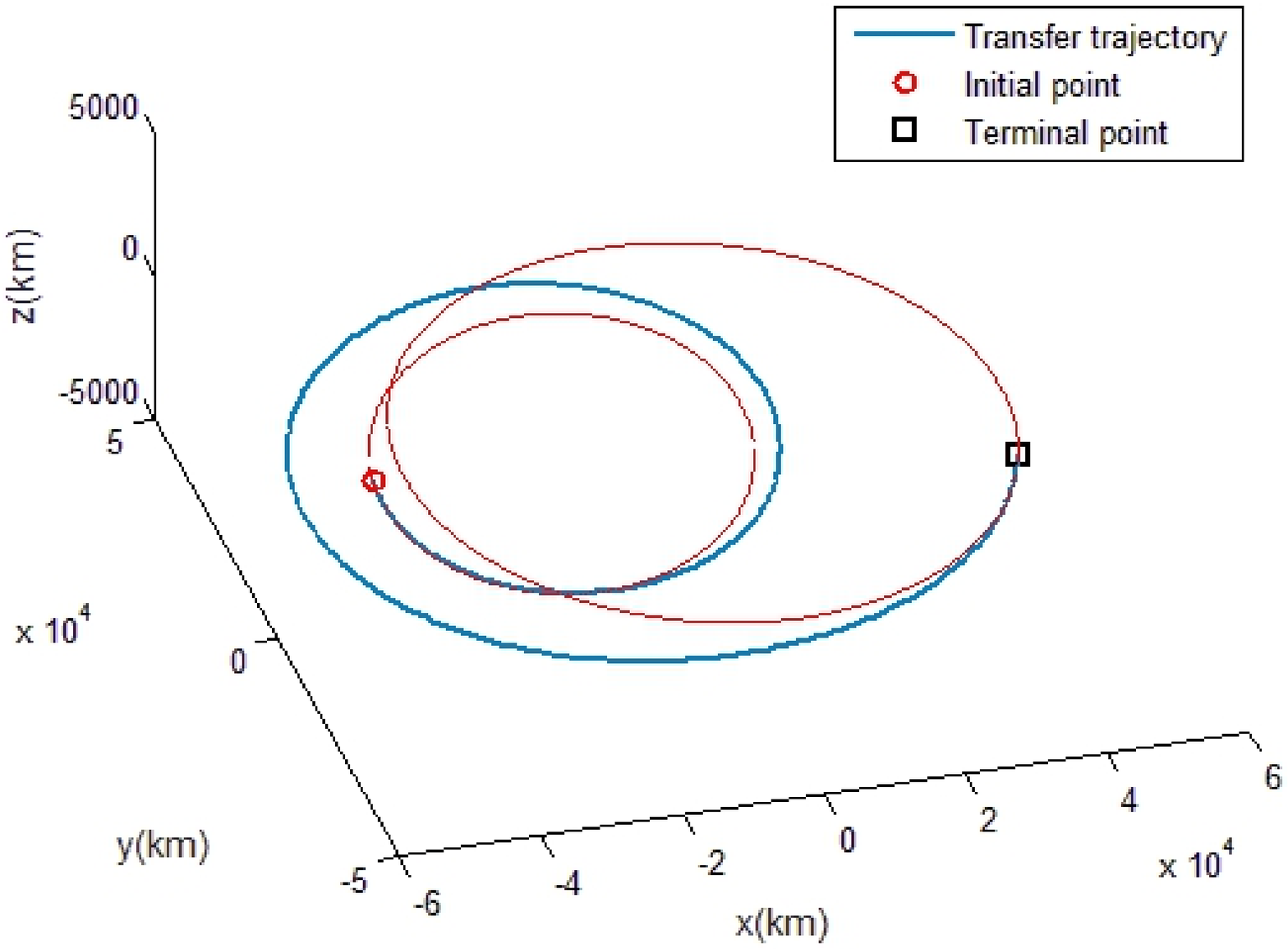}
		\label{Fig:transferT30_a}
	}
	\subfigure[The 85th solution] { \includegraphics[width=0.45\textwidth]{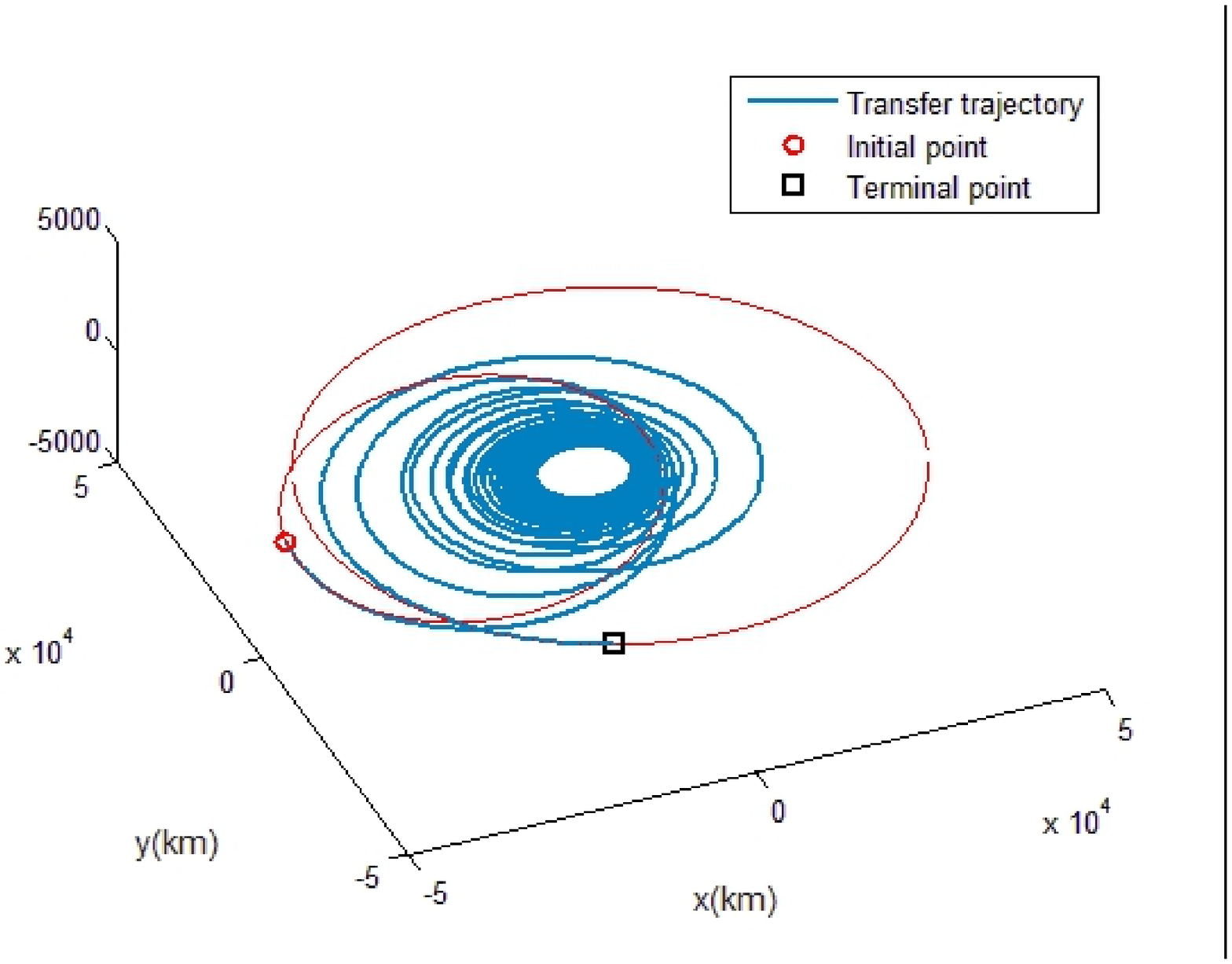}
		\label{Fig:transferT30_b}
	}  
	\caption[]{Three-dimensional minimum-time orbital transfer trajectories for the 1st and 85th solution of the initial problem 
with $\kappa=0$}
	\label{Fig:transfersT30}
\end{figure}

\begin{figure}
	\centering
	\includegraphics[width=0.55\textwidth]{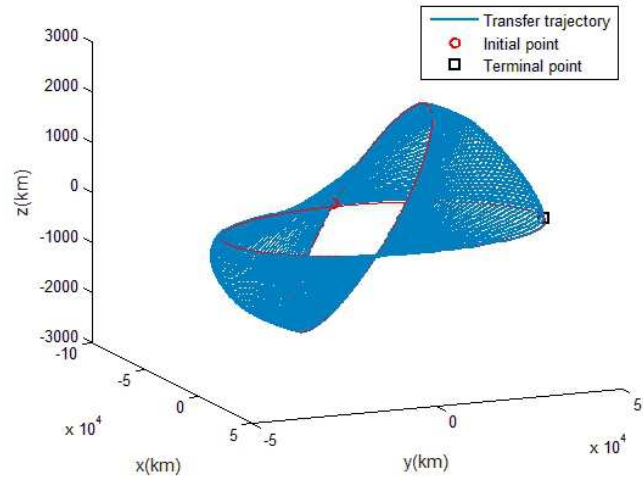}
	\caption[]{Three-dimensional minimum-time orbital transfer trajectory of the original problem with $\kappa=1$}
	\label{Fig:transferT1}
\end{figure}

\begin{figure}
	\centering
	\includegraphics[width=0.9\textwidth]{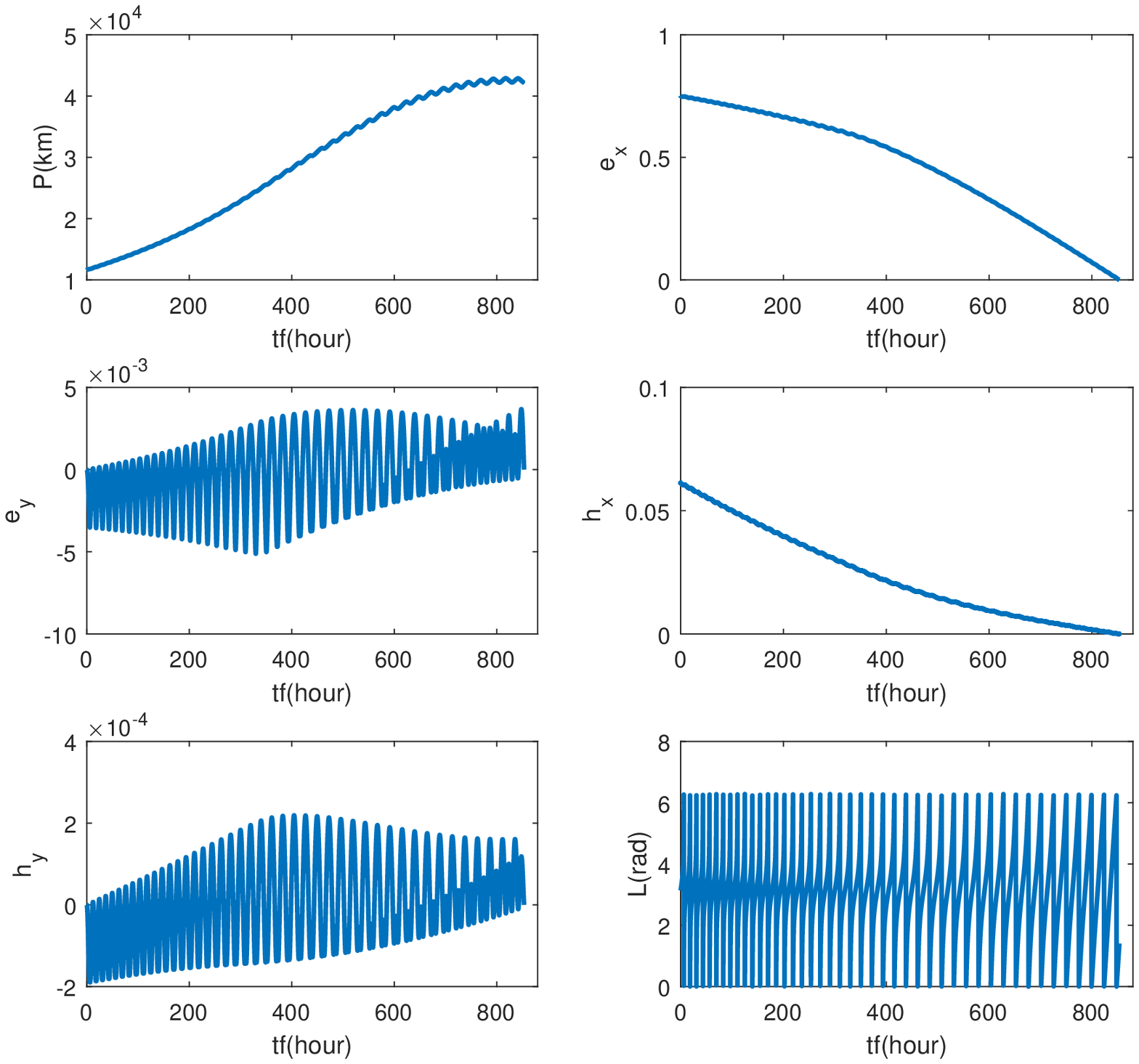}
	\caption[]{Variations of orbit elements along the minimum-time transfer trajectory of the original problem with $\kappa=1$}
	\label{Fig:Elements2time}
\end{figure}

 By the probability-one homotopy method, the continuous zero curve has a total of 85 intersections with $\kappa=0$, the 1st and 85th solution of which, sorted by the obtained terminal time, are provided in Fig.~\ref{Fig:transfersT30}. The three-dimensional minimum-time low-thrust transfer trajectory of the original problem is illustrated in Fig.~\ref{Fig:transferT1}, the transfer time of which is about 853 hours. The variations of the modified equinoctial orbit elements along the minimum-time transfer trajectory are illustrated in Fig.~\ref{Fig:Elements2time}.

\par
\begin{figure}
	\centering
	\subfigure[The complete curve] { \includegraphics[width=0.45\textwidth]{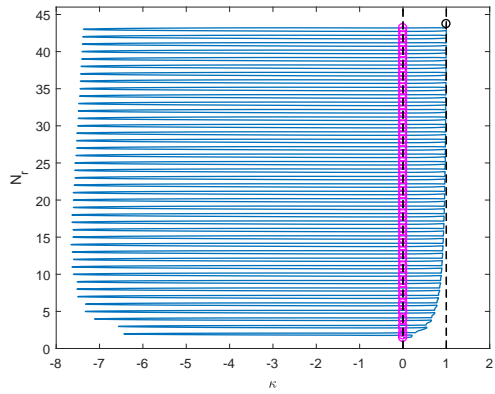}
		\label{Fig:nc2kappa_a}
	}
	\subfigure[The zoom-in view near $\kappa=1$] { \includegraphics[width=0.45\textwidth]{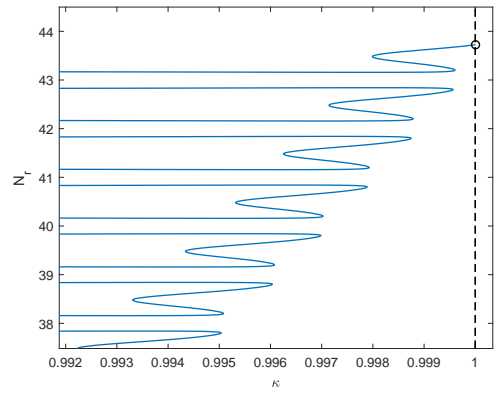}
		\label{Fig:nc2kappa_b}
	}
	\caption[]{Zero curve of $N_r$ by the probability-one homotopy method}
	\label{Fig:nc2kappa}
\end{figure}

 For each value of $\kappa$, define a parameter $N_r$ by
\begin{equation}
N_r = n + \frac{L(t_f)-L_0}{2\pi}
\end{equation} 
where $n$ is the number of the orbital revolutions, $L(t_f)$ is the terminal true longitude, and $L_0$ is the initial true longitude given in Table~\ref{tab:orbit}. If two trajectories have the same integer part in $N_r$,  they have the same number of revolutions along the trajectories. For instance $N_r = 1.1 $ and $N_r  = 1.5$ indicate the same orbital revolutions of 1. Figure~\ref{Fig:nc2kappa} shows the variations of $N_r$ versus $\kappa$ along the zero curve. It should be noted that the revolution numbers of the starting solution and the optimal solution of the original problem are 1 and 43 respectively, as depicted in Fig.~\ref{Fig:nc2kappa}. In Refs.~\citep{Caillau2003,Yue2010Indirect,Pan2016Double}, it is observed that zero curve exists only when the optimal solutions of the initial and the original problem share the same revolution number by their homotopy methods. However, it is demonstrated in this paper that zero curve actually exists even without the the same revolution number requirement by the proposed probability-one homotopy.

\clearpage
\section{Conclusions}
\label{Section:CONCLUSION}
\par This paper presents a new probability-one homotopy method specifically for solving minimum-time low-thrust orbital transfer problems. The generalized sufficient conditions derived in this paper, reasonably explains the failure of the existing homotopy methods, in which the 4th prerequisite of the sufficient conditions is not satisfied, and also ensures the success of the proposed method. The optimal solution of the original problem can be easily found with probability one by tracing the continuous zero curve constructed by the proposed homotopy method. A new discovery is that the continuous zero curve exists even when the initial and the original problem have different number of orbital revolutions, which is not previously known in the literatures. Besides, in the literatures, it was concluded that the homotopy methods are only valid when the initial and the original problem share the same revolution number, which is also incorrect as pointed out in this paper. As illustrated in the numerical demonstrations, the optimal solution of the original problem with 43 revolutions, can be solved by starting from an initial problem with 1 revolution by the proposed method. Thus, the proposed homotopy method provides an efficient approach to find the optimal solution for the minimum-time low-thrust trajectory optimization problems, the convergence of which is probability one.  

\vspace{0.1in}
\begin{center}
{\bf  Acknowledgments }
\end{center}
\par The authors gratefully acknowledge the support  to this work by the National Natural Science Foundation of China (Grant No. 11672234).

\vspace{0.2in}
\clearpage
\bibliographystyle{spr-mp-nameyear-cnd}
\bibliography{bibtex}

\end{document}